\documentclass[aps,pra,twocolumn,superscriptaddress,showpacs]{revtex4-1}
\usepackage{graphicx}
\usepackage{color}

\usepackage[utf8]{inputenc}
\usepackage[english]{babel}
\usepackage{graphicx,tipa}
\usepackage{amsmath}
\usepackage{amssymb}
\usepackage{xcolor}

\renewcommand{\O}[2][]{\hat{#2} ^{\vphantom{\dagger}}_{#1}}
\newcommand{\Op}[2][]{\hat{#2} ^{\dagger}_{#1}}

\usepackage{bm}         

\begin{document}
\bibliographystyle{apsrev}
\title[]{Ultracold Bosons with 
cavity-mediated long-range interactions:\\
A local mean field analysis of the phase diagram
}

\author{Astrid E. Niederle, Giovanna Morigi, and Heiko Rieger}
\email{h.rieger@physik.uni-saarland.de}
\affiliation{Theoretical Physics, Saarland University, Campus E2.6, D--66123 Saarbr\"ucken, Germany}
\date{\today }

\begin{abstract}
Ultracold bosonic atoms in optical lattices self-organize into a variety of
structural and quantum phases when placed into a single-mode cavity and pumped
by a laser. Cavity optomechanical effects induce an atom density modulation at the cavity-mode wave length that
competes with the optical lattice arrangement. Simultaneously short-range interactions
via particle hopping promote superfluid order, such that a variety of
structural and quantum coherent phases can occur. We analyze the
emerging phase diagram in two dimensions by means of an extended Bose-Hubbard model 
using a local mean field approach combined with a superfluid 
cluster analysis. For commensurate ratios of the cavity and external lattice wave lengths the Mott insulator-superfluid transition is modified by the appearance of
charge density wave and supersolid phases, at which the atomic density supports the buildup of a cavity field. 
For incommensurate ratios, the optomechanical forces induce the formation of Bose-glass and superglass phases, namely non-superfluid and superfluid phases, respectively, displaying quasi-periodic density modulations, which in addition can exhibit structural and superfluid stripe formation. The onset of such structures is constrained by the onsite interaction and is favourable at fractional densities. Experimental observables are identified and discussed.
\end{abstract}


\maketitle

\section{Introduction}

The quantum phase transition between the Mott-Insulator (MI) and Superfluid (SF) phase of the Bose-Hubbard model is a paradigmatic example of strongly correlated systems which can be realised with ultracold atomic ensembles in optical lattices \cite{Fisher:1989,Sachdev:Book,Bloch:2010}. This dynamics  results from the interplay between the short-range nearest-neighbor hopping, which  promotes delocalization and superfluidity and can be controlled by the lattice depth, and the on-site repulsion, which penalizes high local densities and can be tuned by means of Feshbach resonances \cite{Jaksch:1999,Bloch:2010}. 

\begin{figure}[htbp]
\begin{center}
 \includegraphics[width=0.4\textwidth]{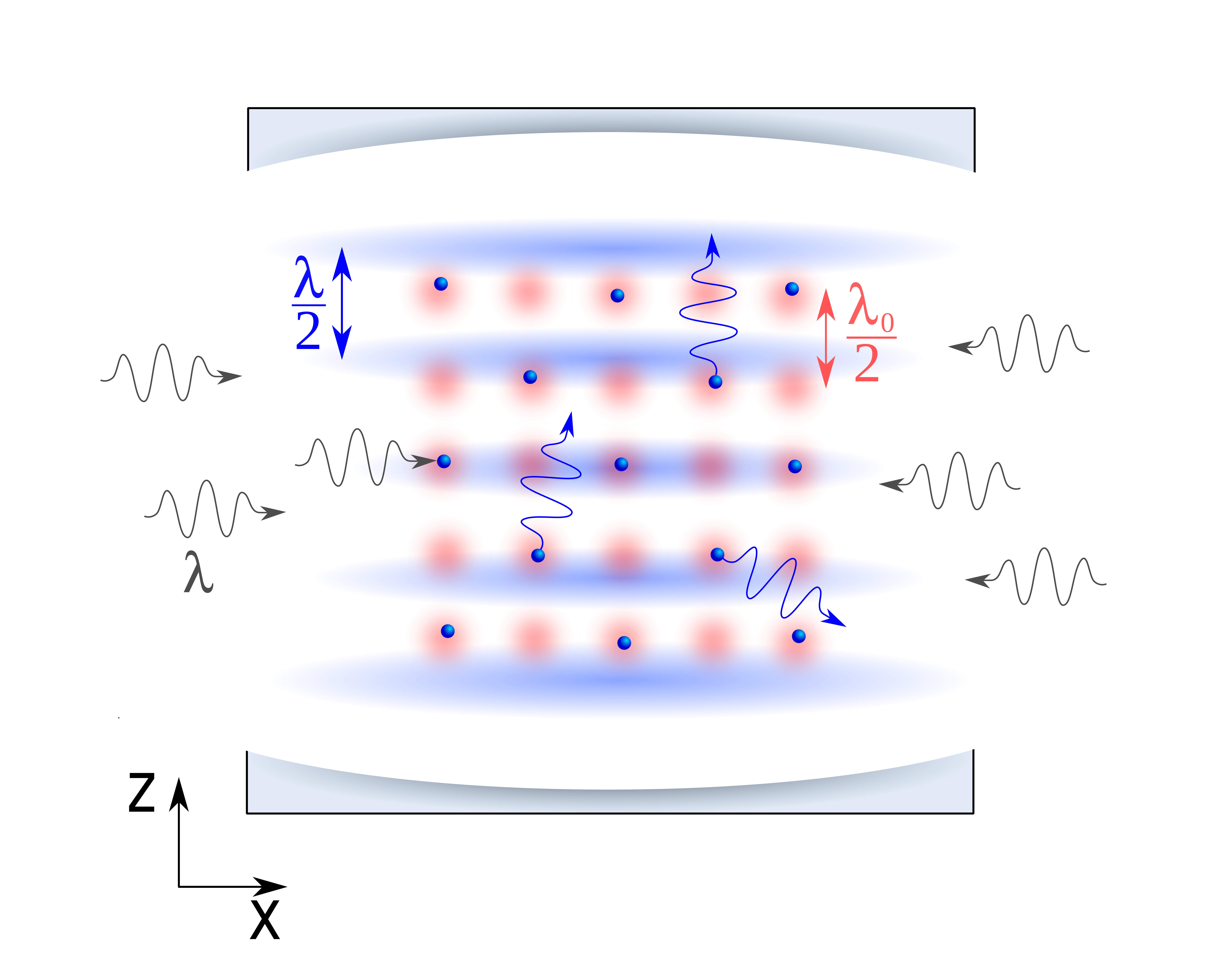}
\caption{\label{Fig:1}Bosonic atoms are confined by a two-dimensional tight optical lattice at wave length $\lambda_0$ within a high-finesse cavity and dispersively interact with a standing-wave mode at wavelength $\lambda$. The cavity mode is lossy and pumped by coherent scattering of the atoms, which are transversally pumped by a laser with the same cavity wave length $\lambda$. The atoms quantum phases result from the interplay between the kinetic energy, the onsite $s$-wave scattering, and the long range interaction induced by the cavity forces. In this work we apply a local mean field approach with a cluster analysis in order to determine the structural and superfluid order of the emerging phases for different values of the ratio  $\lambda_0/\lambda$.}
\end{center}
\end{figure}

If additional interactions are added, whose length scales compete with the periodicity of the optical lattice, further phases can appear where superfluidity and periodic (or quasi-periodic) structural order coexist \cite{Frey:1997,Otterlo:1994,Batr:1995,Scal:1995,Otterlo:1995,Fazio:2011,Baumann:2010,Nagy:2010,Ritsch:2013}. A prominent example is the long-range interaction mediated by photon scattering of atoms in a high-finesse resonator \cite{Nagy:2010,Baumann:2010,Ritsch:2013}: Here, when the atoms are transversally driven by a laser and coherently scatter photons into the cavity, as illustrated in Fig. \ref{Fig:1}, then the mechanical effects of light on the atoms can give rise to a periodic potential with the periodicity of the cavity wavelength $\lambda$ and whose depth is a function of the atomic density itself \cite{Asboth:2005,Fernandez:2010,Keeling:2010,Bhaseen:2010}. Recent works analysed how this long-range interaction modifies the phase diagram of ultracold bosons when it is a weak perturbation to the external periodic potential \cite{Fernandez:2010,Habibian:2013,Li:2013,Reza-Baktiari:2015,Klinder:2015} and when its strength competes with the onsite repulsion \cite{Landig:2015,Zhai:2016,Dogra:2016,Caballero:2016}, demonstrating the existence of novel phases which are associated with structural order sustaining coherent scattering into the resonator.  In Ref. \cite{Habibian:2013}, in particular, some of us analysed the phase diagram emerging when the wavelength of the external optical lattice $\lambda_0$ is incommensurate with the cavity wave length $\lambda$ for the setup of Fig. \ref{Fig:1}, assuming that the atoms are tightly-bound at the minima of the external potential. For this configuration a Quantum Monte-Carlo simulation in one dimension predicted the appearance of compressible phases, with vanishing superfluidity, which we denoted by Bose-glass (BG) phase using the terminology applied for cold atoms in bichromatic, aperiodic potentials \cite{Orignac,Roux,Deissler}. Differing from the situations considered in Refs. \cite{Deissler,Orignac,Roux}, however, here the second potential is created by atomic scattering and can exist only if there is some type of Bragg order. Indeed, we found that in the BG phase the atomic density is modulated at the beating wave length of the two potentials, thus supporting coherent scattering into the resonator \cite{Habibian:2013}. In two dimensions, a mean field calculation predicted that such structural order can also coexist with superfluidity \cite{Habibian:2013}. Nevertheless, the type of mean field approach there applied does not allow to precisely determine the boundary of the transition from BG to SF phase, and thus only delivers a qualitative picture of the phase diagram.

In the present work we perform a systematic characterization of the two-dimensional phase diagram for the setup in Fig. \ref{Fig:1}, which is based on a local mean field (LMF) approach with cluster analysis, developed by some of us in Ref. \cite{Niederle:2013} for the disordered Bose-Hubbard model. This study allows us to get a quantitative insight into the interplay between superfluidity and long-range interactions.
We determine the phase diagram for commensurate and incommensurate values of the ratio $\lambda_0/\lambda$, comparing in particular the commensurate case $\lambda=\lambda_0$ with the incommensurate one $\lambda=\lambda_0(1+\epsilon)$, when $\epsilon$ is a small irrational number. We provide a classification of the phases we identify and extract a detailed picture of the onset of structural order induced by cavity backaction. 

This article is organised as follows. In Sec. \ref{Sec:Model} we introduce the extended Bose-Hubbard model, which is at the basis of our study, and the local mean field approach with cluster analysis. In Sec. \ref{Sec:C} we discuss the phase diagram when the lattice and cavity wavelength are commensurate, while in Sec. \ref{Sec:I} we focus on the incommensurate case. The conclusions are drawn in Sec. \ref{Sec:Conclusions}.

\section{Extended Bose-Hubbard model}
\label{Sec:Model}

The quantum phases of the ultracold bosons in the setup of Fig. \ref{Fig:1} are found starting from the effective Hamiltonian derived by means of a coarse-graining procedure \cite{Larson:2008,Habibian:2013}. The Hamiltonian  describes the dynamics of an ultracold gas of atoms confined to the $x-z$ plane, which are tightly confined by an external square lattice with lattice constant $a=\lambda_0/2$ and which interact via $s$-wave collisions and through the long-range cavity forces. These forces emerge from the adiabatic part of the cavity dynamics, under the assumption that the typical time scale of the cavity field evolution is much shorter than the one characterizing the atomic motion. The Hamiltonian is cast in the form of a Bose-Hubbard model,  assuming that the dynamics is restricted to the lowest band of the external lattice and that the interaction with the resonator is sufficiently weak so that interband transitions are suppressed. We denote by $w_i(x,z)$ the Wannier functions for the optical lattice lowest band. Here, $i=(i_x,i_y)$ labels the site and $i_x,i_y=1,\ldots,L$, such that the size is $K=L^2$ with periodic boundary conditions.

Below, we adopt the convention that the cavity axis is along $z$ and that the cavity spatial mode function is $\cos(kz)$, with $k=2\pi/\lambda$. The atoms are pumped by a standing-wave laser propagating along $x$ with the same wave number $k$, forming an optical lattice with intensity distribution $\propto  \cos^2(k x)$. The laser frequency $\omega_L$ is detuned by $\Delta_c=\omega_L-\omega_c$ from the cavity-mode frequency $\omega_c$. The sign of this detuning determines whether Bragg  gratings are energetically favourable \cite{Asboth:2005,Domokos:2002,Schuetz:2013,Piazza:2015}. In this work we restrict our analysis to the case $\Delta_c<0$, for which spatial selforganization can occur.

\subsection{Effective Hamiltonian}

The grand-canonical Hamiltonian $\hat{H}$, which is the starting point of our analysis, is reported in second quantization in terms of the bosonic creation and
annihilation operators of an atom at site $i$, denoted by $\hat{a}^\dagger_i$ and $\hat{a}_i$, respectively, such that $[\O[i]{a},\Op[j]{a} ]=\delta_{i,j}$, and takes the form \cite{Habibian:2013}
\begin{eqnarray}\label{eq:BH}
\hat{H}&=&-J \sum_{\langle i,j \rangle} \left(\Op[i]{a} \O[j]{a}+\O[i]{a} \Op[j]{a}\right)+ \frac{U}{2} \sum_{i} \O[i]{n} \left( \O[i]{n}-1\right) 
\nonumber\\
            &  & +\sum_{i} (\epsilon_i-\mu) \O[i]{n}
            + K\hat \delta\hat \Phi^2\,,
\end{eqnarray}
where $\O[i]{n}=\Op[i]{a} \O[i]{a}$ is the local number operator at site $i$. 
The first three summands on the right hand side (RHS) of Eq. \eqref{eq:BH} are (i) the kinetic energy, scaled by the hopping strength $J$, (ii) the onsite collisions
due to $s$-wave scattering with $U$ the strength of the interaction, and (iii) the onsite energy  $\epsilon_i$ and the chemical potential $\mu$. 
The onsite energy $\epsilon_i=\epsilon_0+V_1Y_i ^{(x)}$ has a constant offset 
$\epsilon_0$ and a site dependence due to the transverse laser potential which pumps the atoms: Here, $V_1$ is the depth of the transverse laser optical lattice at wavelength $\lambda$ and the shift it induces at the sites $i$ is given by
\begin{equation}
\label{Y:i}
Y_i^{(x)}= \int dx\int dz\, w_i^2(x,z) \cos^2(k x)\,.
\end{equation}
Hence, in absence of the cavity field the atoms experience a bichromatic potential along the $x$ axis, while along $z$ the potential is periodic with periodicity $a$. 

The last term of the Hamiltonian represents the long-range interaction mediated by the cavity field, where $\hat \delta$ is an operator depending on the atomic density distribution, which converges to a finite value in the thermodynamic limit we apply and whose specific form is given at the end of this Section. 
Finally,
\begin{equation}
\hat \Phi=\frac{1}{K}\sum_i Z_i\hat{n}_i
\end{equation} 
depends on the density at each site $i$, and it is weighted by the scattering amplitudes $Z_i$ at site $i$. The scattering amplitudes read
\begin{equation}
\label{Z:i}
Z_i= \int dx\int dz \, w_i^2(x,z)\cos(kx)\cos(kz)\,,
\end{equation} 
and are $\lambda$-periodic numbers. Thus, $|\langle \hat \Phi\rangle|\le \bar n$, with $\bar n$ the average onsite density.

For the parameters we choose the sign of ${\hat \delta}$, and thus of the potential, 
is negative. The cavity long-range interactions thus favour configurations where the expectation value of the global operator $\langle \hat \Phi^2\rangle$ is maximized. This expectation value, in turn, is proportional to
the mean intracavity photon number and is maximum when the atom density modulation has period $\lambda$, thus forming a Bragg grating.
In fact, we show below that when $\langle \hat\Phi^2\rangle\neq 0$, the structure form factor exhibits a peak at the cavity wave vector.

\subsection{Local mean field with superfluid cluster analysis}

We analyze the phase diagram emerging from the Hamiltonian (\ref{eq:BH})
using local mean field (LMF) theory  \cite{shes93} combined with a superfluid cluster analysis \cite{Niederle:2013,Niederle:2015}. Within this framework, we define the 
so-called local SF parameter 
by equation $\psi_i=\langle\O[i]{a} \rangle$, which has to be determined self-consistently, being the expectation value of operator $\O[i]{a}$ taken over the ground state of the LMF Hamiltonian $\hat{H}_\mu^{\rm MF}=\sum_i \hat H_i^{\rm MF}$, with
\begin{eqnarray}\label{HLMF}
 \hat H_i^{\rm MF}&=&- \eta_i\,J  \left(\Op[i]{a} +\O[i]{a}-\psi_i\right)+ \frac{U}{2} \O[i]{n} (\O[i]{n}-1) \nonumber\\
                             & &+(\epsilon_i- \mu) \O[i]{n}+\langle{\hat \delta}\rangle Z_i\langle\hat \Phi\rangle \hat n_i \, , 
\end{eqnarray}
and $\eta_i=\sum_{j n. n.} \psi_j$ is the sum of the SF parameters $\psi_j$ at the nearest neighbors of site $i$, and $\langle{\hat \delta}\rangle$, $\langle\hat \Phi\rangle$ are determined self-consistently. 

The sites are classified depending on the value of the onsite particle number fluctuations, namely, $$\Delta n_i^2=\langle \hat n_i^2\rangle-\langle \Delta \hat n_i\rangle^2\,.$$ The sites with $\Delta n_i^2=0$ are denoted by Mott-Insulator (MI) sites, otherwise they are SF sites. We denote by $N_{SF}$ the number of the SF sites and by $P_j=1$ the existence of a percolating line connecting two opposite sides of the lattice along $j=x,z$, otherwise it is zero. Using these quantities we determine the phase. 

We first review the phases which typically emerge from the interplay of kinetic energy and onsite interaction, and in absence of the cavity potential. 
A phase is incompressible provided that all sites are MI, thus $N_{SF}=0$. A BG phase exhibits clusters of SF sites, that are surrounded by MI sites: Within a cluster of SF sites particle are allowed to tunnel freely, which means that these sites form a local SF island in a background of MI sites. The phase is SF$_j$, namely, SF along the $j=x,z$ direction, if at least one of these clusters can percolate in the $j$ direction. When $P_x=P_z=1$, the phase is SF.  

The cavity potential, on the other hand, favours a cavity-mediated long-range order whose signature is the nonvanishing expectation value of operator $\hat \Phi$. The onset of this order can modify the phase, that otherwise would characterize the atoms ground state. Below we discuss commensurate (C) and incommensurate (I) configurations, for which the MI or the normal SF do not support the presence of a stationary cavity field, being $\sum_i Z_i=0$ and thus $\langle \hat \Phi\rangle=0$ \cite{Habibian:2013}. The corresponding classification is summarized in Table \ref{Table:1}. 

We now argue that for $\lambda=\lambda_0$ the cavity-mediated potential tends to induce a $(\pi,\pi)$ 
density modulation, also denoted as charge density wave (CDW), in the form of a checkerboard pattern with alternating site occupation numbers. For this purpose we introduce the structure form factor operator, which is defined as 
\begin{equation}
\hat S(k_x,k_z)=\frac{1}{K}\sum_{i,j}
\exp(-i(k_xj_x+k_zj_z)a) \hat n_i\hat n_j
\end{equation}
Using Eq. \eqref{Z:i}, and that for this commensurate case $Z_i\propto(-1)^i$, then
\begin{equation}
\hat S(\pi,\pi)=
K\hat \Phi^2\;.
\end{equation}
Thus, in addition to the homogeneous phases, SF and MI, also a CDW long-range order with $\langle\hat \Phi\rangle\ne0$ can occur. This diagonal long-range 
order can in principle also coexist with off-diagonal SF order, which is then commonly denoted as a supersolid (SS) phase \cite{Otterlo:1994,Batr:1995,Scal:1995,Otterlo:1995,Fazio:2011}. We will denote the phase supersolid in $j=x,z$ direction (SS$_j$) when the phase is SF only along $j$ (namely, when $P_j=1$ but vanishes in the other direction). Correspondingly, when instead the ratio is incommensurate, the density modulation is quasi-periodic. The phase is a BG when $N_{SF}=0$ and a Super-Glass in $j=x,z$ direction (SG$_j$) if it coexists with SF order.

\begin{table}
\begin{center}
 \begin{tabular}{|l|c|c|c|c|c|l|c|c|c|c|c|}
\hline
{\rm Phase C}&$N_{SF}$& $P_\text{X}$& $P_\text{Z}$&$\langle\O{\Phi}\rangle$&&{\rm Phase I}&$N_{SF}$& $P_\text{X}$& $P_\text{Z}$&$\langle\O{\Phi}\rangle$\\ \hline
MI&$ 0$&$ 0$&$0$&$0$&&MI&$ 0$&$ 0$&$0$&$0$\\ \hline
CDW&$0$&$0$&$0$&${\rm yes}$&&BG&${\rm yes}$&$0$&$0$&${\rm yes}$\\ \hline
SF&${\rm yes}$&$1$&$1$&$0$&&SF&${\rm yes}$&$1$&$1$&$0$\\ \hline
SF$_\text{X}$&${\rm yes}$&$1$&$0$&$0$&&SF$_\text{X}$&${\rm yes}$&$1$&$0$&$0$\\ \hline
SF$_\text{Z}$&${\rm yes}$&$0$&$1$&$0$&&SF$_\text{Z}$&${\rm yes}$&$0$&$1$&$0$\\ \hline
SS&${\rm yes}$&$1$&$1$&${\rm yes}$&&SG&${\rm yes}$&$1$&$1$&${\rm yes}$\\ \hline
SS$_\text{X}$&${\rm yes}$&$1$&$0$&${\rm yes}$&&SG$_\text{X}$&${\rm yes}$&$1$&$0$&${\rm yes}$\\ \hline
SS$_\text{Z}$&${\rm yes}$&$0$&$1$&${\rm yes}$ &&SG$_\text{Z}$&${\rm yes}$&$0$&$1$&${\rm yes}$  \\ \hline
\end{tabular}
\end{center}
\caption{\label{Table:1} Observables determining the phases for the commensurate (C, left) and, correspondingly, for the incommensurate (I, right) quantum potential. $N_{SF}$ indicates whether the number of SF sites is larger than zero; $P_\text{X,Z}=1$ when percolating lines, signalling superfluidity, exist along the $x,z$ directions. The expectation value $\langle\O{\Phi}\rangle\neq 0$ is found when the atomic density forms a Bragg grating, corresponding to a peak of the structure form factor at the cavity wave vector along the $z$ axis with wave number $k$. }
\end{table}

\subsection{Experimental parameters}
\label{Sec:parameters}

The Hamiltonian we consider describes the coherent dynamics of a driven-dissipative system, which emerges within a coarse-grained description. The energy of the atoms external degrees of freedom is thus conserved when the cavity field rapidly relaxes to a state that is determined by the atomic density distribution \cite{Larson:2008,Habibian:2013}. The parameters of the long-range interactions depend on the characteristic parameters of the photon scattering dynamics, namely, on (i) the coherent scattering amplitude $S_0$ that determines the rate at which the cavity is pumped. It reads  $S_0=\Omega g /\Delta_a$ where $\Omega$ is the laser Rabi frequency, $g$ the vacuum Rabi frequency, and $\Delta_a$ the detuning of the fields from the atomic dipole transition; (ii) the cavity loss rate $\kappa$ whose interplay with $S_0$ gives the mean intracavity photon number; (iii) the bare detuning between laser and cavity field $\Delta_c$, such that the frequency offset between laser reads 
\begin{equation}
{\hat \delta}_{\text{eff}}=\Delta_c-U_0\sum_i Y_i ^{(z)}\hat n_i\,, 
\end{equation}
where the second term on the right-hand side is the dynamical Stark shift due to the atomic density distribution with $U_0=g^2/\Delta_a$. Moreover, the depth of the transverse laser optical lattice is $V_1=\hbar\Omega^2/\Delta_a$ and it is connected to the other two characteristic frequencies $S_0$ and $U_0$ by the relation $V_1=\hbar S_0^2/U_0$.
In this work we rescale these quantities according to a thermodynamic limit of Refs. \cite{Asboth:2005,Fernandez:2010}, where $U_0=u_0/K$ and $S_0=s_0/\sqrt{K}$. In particular, the long-range potential in Eq. \eqref{eq:BH} reads 
 \begin{equation}
\label{delta}
 \hat \delta =\frac{\hbar s_0^2}{{\hat \delta}_{\text{eff}}^2+\kappa^2} {\hat \delta}_{\text{eff}}\,.
 \end{equation}
Our description is valid when the atoms are in the lowest band of the external optical lattice, so that the long-range interaction is a perturbation. We will use here the parameters for the two-dimensional calculation in Ref. \cite{Habibian:2013}, namely: $s_0=0.15\kappa$, $u_0=237\kappa$, $\Delta_c=-0.5\kappa$ and $\kappa=2\pi\times 1.3$ MHz which is consistent with the cavity setup of Ref. \cite{Baumann:2010}. The strength of the onsite potential is found by performing the integral $U=g_0\int dx \int dz w(x,z)^4$ with $g_0/\hbar=5.5\times 10^{-11}$ Hz m$^2$ \cite{Kruger:KT}.
Consistent with the experimental setup of Ref. \cite{Baumann:2010}, we choose $\lambda_0=\lambda=830$ nm while, for the incommensurate case $\lambda=785$ nm and $\lambda_0=830$ nm. 

\section{Commensurate wave lengths }
\label{Sec:C}

We first consider the phase diagram when the optical lattice and the cavity mode have commensurate wave length.
We focus on the case $\lambda=\lambda_0$, which has been recently experimentally realised, and shortly discuss other commensurate ratios at the end of this section.

\subsection{Preliminary considerations}

To better understand which phases 
could actually occur in the present model we inspect a slightly simplified Hamiltonian, in which we neglect the
site-dependence of the on-site potential $\epsilon_i$ (i.e. setting 
$\epsilon_i=0$), and which corresponds to the models considered in Refs. \cite{Li:2013,Reza-Baktiari:2015,Zhai:2016,Dogra:2016}. We further perform the substitution $\langle\hat \delta\rangle\to -\delta$ in Eq. \eqref{HLMF}, where $\delta>0$.
Then
\begin{eqnarray}
\hat{H}'&=&-J\sum_{\langle i,j \rangle} \left(\Op[i]{a} \O[j]{a}+\O[i]{a} \Op[j]{a}\right)+ \frac{U}{2} \sum_{i} \O[i]{n} \left( \O[i]{n}-1\right) 
\nonumber\\
&  & -\mu\sum_{i} \O[i]{n} 
- \frac{\delta}{K}\Bigl(\sum_{i,even}\hat n_i - \sum_{i,odd}\hat n_i\Bigr)^2\,,
\label{Hsimp}
\end{eqnarray}
where $\sum_{i,even}$ ($\sum_{i,odd}$) is restricted to the sites where $i_x$ and $i_z$ are even (odd) numbers. 
If the cavity coupling $\delta$ is larger than the on-site repulsion $U$, $\delta>U$, a complete even/odd
imbalance (i.e.\ either all even sites or all odd sites
empty) is energetically favorable. 

The phase diagram as a function of the ratio $\delta/U$ has
been evaluated in Ref. \cite{Dogra:2016} by means of a mean field treatment, supported by results from quantum Monte Carlo simulations for the hard-core ($U\to\infty$) limit. Here, we make some general considerations on the basis of analogies with a well known model. 
These considerations start from the observation that Hamiltonian $\hat{H}'$ is invariant under all discrete translations that leave this checkerboard pattern invariant. Therefore,
its ground state must possess the same symmetry, which implies that all even sites and all odd sites have identical properties 
in the ground state. Consequently, the mean field approximation in Eq. (\ref{HLMF}) for $\hat H'$ leads to an effective two-site Hamiltonian of the form
$\hat H_{\rm MF}'=\frac{K}{2}(\hat h_e+\hat h_o)$, where  $\hat h_e$ and $\hat h_o$ refer to the single site MF Hamiltonians for even and odd sites, respectively, and read 
\begin{eqnarray}
\hat h_\ell&=&-J\eta_\ell(\hat a_\ell^\dagger+\hat a_\ell-\psi_\ell)+ \frac{U}{2}\,\hat n_\ell(\hat n_\ell-1)\nonumber\\
&&-\mu \hat n_\ell -\frac{\delta}{2} \hat n_\ell^2
+\frac{\delta}{2} \hat n_e \hat n_o\;,
\label{HLMF2}
\end{eqnarray}
where $\ell=e,o$, $\eta_e=4\psi_o$, and $\eta_o=4\psi_e$. It is interesting to note that the last term on the right-hand side of Eq. \eqref{HLMF2}, $(\delta/2)\,\hat n_e \hat n_o$, has the form of a 
repulsive nearest-neighbor interaction between bosons on neighboring even and odd sites and is also characteristic for the mean field theory of the extended Bose-Hubbard model 
with nearest-neighbor repulsion, where the interaction takes the form $V \sum_ {\langle i,j \rangle} \hat n_i \hat n_j$, see Refs. \cite{Batr:1995,Scal:1995,Otterlo:1995,Iskin,Kimura}.
Here, it was shown that when the nearest-neighbor  interaction $V$ is sufficiently strong, namely, of the same order of magnitude as the on-site 
repulsion $U$, then it can stabilize a CDW order even inside the SF phase, yielding alternating MI and CDW insulating lobes whose centers are shifted by half integer values of $\mu/U$. This model further predicts that the CDW insulating lobes are equipped with a supersolid SS tip regions. On the basis of the equality of the mean field theories for Eq. \eqref{Hsimp} and for the extended Bose-Hubbard model their mean field phase diagrams for Eq. \eqref{Hsimp} look as those depicted in Ref. \cite{Iskin}. Nevertheless,  in Eq. \eqref{HLMF2} the effective nearest neighbor interactions
strength $\delta$ diminishes the on-site repulsion. For $\delta>U$, in particular, the effective onsite interaction becomes attractive, for which reason the grand-canonical ensemble becomes invalid. For fixed particle densities, at small $J$ the phase is CDW with either the even or the odd sites occupied. 

Further insight can be gained in the limit of strong on-site repulsion ($U\to\infty$), denoted as the 
limit of hard-core bosons. In this limit $n_\ell$ can only be 0 or 1. Then via the identification 
$\hat a_i^\dagger=\hat S_i^\dagger=\hat S_i^x+i\hat S_i^y$ and $\hat n_i=\hat S_i^z+1/2$ ($\hat S_i^x$, $\hat S_i^y$, $\hat S_i^z$ the spin-1/2 operator) the Hamiltonian (\ref{Hsimp}) maps onto the spin-1/2 Hamiltonian
\begin{eqnarray}
\hat{H}_{\rm spin}
&=&-t\sum_{\langle i,j \rangle} \left(\hat S_i^x\hat S_j^x+\hat S_i^y\hat S_j^y\right)
-h\sum_i \hat S_i^z 
-\frac{\delta}{L^2} \hat M_{\rm stagg}^2,\nonumber\\
& &
\label{Hspin}
\end{eqnarray}
where $h=\mu-\delta/2$, $t=2J$, and
$\hat M_{\rm stagg}=(\sum_{i,even} \hat S_i^z -\sum_{i,odd} \hat S_i^z)$
is the total staggered magnetization of the square lattice and thus
the order parameter for N\'eel (i.e. antiferromagnetic) long-range order 
of the spin system. Thus, in the hard-core limit the cavity mediated 
long-range interaction is identical to a mean field and thus maximally 
long-ranged antiferromagnetic (AFM) interaction among the spins $z$ component.
The hopping term in (\ref{Hsimp}) on the other hand translates into a ferromagnetic (FM) short-range interaction in the spins XY plane. 
Thus CDW / SF order in the bosonic system is equivalent to AFM-z / FM-xy 
order in the spin system. A supersolid phase then corresponds to 
simultaneous AFM spin order in the $z$ direction and 
FM spin order in the $xy$ plane. Both types of order can in principle
coexist in magnetic systems, as for instance in the fully symmetric 
Heisenberg model \cite{Baxter}.

\begin{widetext}
\subsection{Phase diagram}

\begin{figure}[!htb]
\begin{center}
\includegraphics[width=0.8\textwidth]{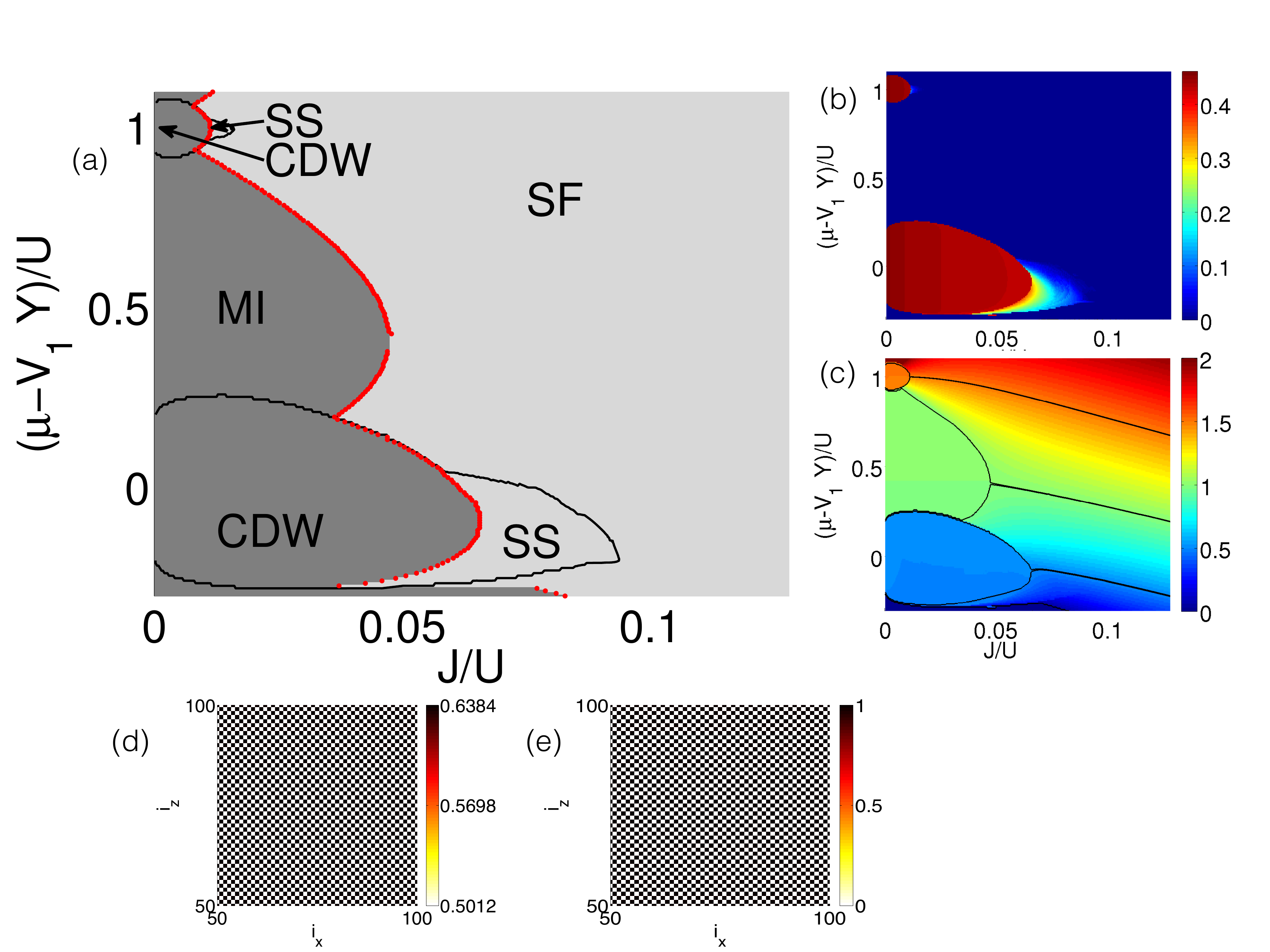}\\
 \caption{\label{PhaseOrdered1} Phase diagram in the $J-\mu$ plane evaluated from Eq. \eqref{HLMF} using LMF with cluster analysis when the atoms are trapped in squared lattice with interparticle distance $a=\lambda_0/2$ and interacting with a cavity field along the $z$ direction with $\lambda=\lambda_0$. Here, the chemical potential is shifted by the onsite energy $V_1Y$, with $Y=Y^i_x$ given in Eq. \eqref{Y:i}. (a) Classification of the phases according to Table \ref{Table:1}. The red lines separate the incompressible from the compressible phases, the black lines the regions with Bragg order from the ones where the intracavity field vanishes (see Table \ref{Table:1}). (b) Contour plot of the expectation value of operator $\hat\Phi$ and (c) of the mean atomic density as a function of $\mu/U$ and $J/U$. The lattice is composed by $K=100\times 100$ sites with periodic boundary conditions. Subplots (d) and (e) display the density distribution in the lattice in the SS and in the CDW phase, respectively. See Sec. \ref{Sec:parameters} for the other parameters.}
 \end{center}
\end{figure}
\end{widetext}

We now analyse the predictions of the LMF with SF cluster analysis for Hamiltonian \eqref{HLMF}. 
Figure \ref{PhaseOrdered1}(a) displays the $\mu-J$ phase diagram for $\lambda=\lambda_0$, where we labeled the phases according to the criteria listed in the left Table \ref{Table:1}. Besides  the MI and the SF phases, we observe the appearance of a Charge-Density Wave (CDW) and a Supersolid (SS) phase at fractional densities. As visible in subplot (b), these phases correspond to a non-vanishing intracavity photon number ($\langle \hat \Phi\rangle \neq 0$), such that the cavity field is maximum in the CDW phase, at sufficiently small hopping rates, and it gradually decreases to zero as $J$ increases (and thus the density modulation decreases) in the SS phase. By changing $\mu$ and/or $J$  the CDW has a direct transition to either a MI, a SF, or a SS phase. 

Inspection of subplot (c), depicting the contour plot of the mean density, shows that the transition between CDW and MI or SF is associated with a jump in the value of the mean density, while the CDW and SS phases occur for the same value of the mean density and in particular for $\bar n=1/2$ and $\bar n=3/2$. Indeed, for the chosen parameters these densities (as well as all other fractional densities $(2\ell+1)/2$ not shown in the plot) allow for the buildup of a Bragg structure (the corresponding Bragg-ordered region decreases as $\ell$ increases since for larger densities the onsite interaction makes them energetically unstable). In the CDW and for $\bar n=1/2$ a filled site is surrounded by empty sites, giving rise to a checkerboard structure. Using the notation introduced in Ref. \cite{Dogra:2016}, the occupation of any pair of nearest neighbours can be symbolized by the vector $(1,0)$, whose entries give the occupations of the two neighbouring sites. For $\bar n=3/2$ the checkerboard structure is instead $(2,1)$. Thus, for the chosen strength of the long-range interacting potential Bragg order can only occur at fractional densities and can coexist with SF off-diagonal order. This result is in agreeement with the supersolidity revealed at the Dicke phase transition \cite{Nagy:2010,Baumann:2010}. It is also interesting to compare this phase diagram with the one obtained when the atoms are solely trapped by the potential they scatter: In this case one also observes a transition from SF to SS, followed by CDW at fractional densities $\bar n=1/2,3/2,...$ Between these phases there is a gap of values of the chemical potential, where the phase remains SS \cite{Fernandez:2010}. Figure \ref{PhaseOrdered1}(a) shows that, in presence of an external commensurate potential and at small $J$, the phases at fractional densities remain CDW while within the gaps the phase is MI. 

We comment that, if the strength of the long-range interaction is increased with respect to the onsite repulsion, thus when $\delta/U$ is sufficiently large, other type of checkerboard structures can appear at commensurate and other fractional densities. For instance, at $\bar n=1$ a Bragg structure can appear with occupation $(2,0)$, for $\bar n=3/2$ checkerboard with the occupation $(3,0)$ \cite{Dogra:2016}.  The parameter choice we have made in this work is such that the onsite repulsion dominates, therefore only the fractional densities $\bar n=(2\ell+1)/2$ favour the formation of stable CDW phases. 

For the same value of $\delta/U$ we have further analysed the phase diagrams for other commensurate ratios $\lambda_0/\lambda$ focussing on the values $\lambda_0/\lambda=\ell$ with $\ell$ integer. We identify two classes of behaviour: For $\ell$ odd the phase diagram has an analogous form to the one of Fig. \ref{PhaseOrdered1}, showing CDW phases with a SS tip at the fractional densities, for which Bragg gratings can form. The value of $\ell$ even is particular, since for this choice the MI is a Bragg grating. Therefore, the phase diagram exhibits MI and SF phases and all over the phase diagram the number of intracavity photons is different from zero: it is larger at small tunneling, when the density modulation is larger, and it increases with the onsite density, thus with $\mu$.

\begin{widetext}

\section{Incommensurate wave lengths}
\label{Sec:I}

\begin{figure}[!htb]
\begin{center}
\includegraphics[width=0.8\textwidth]{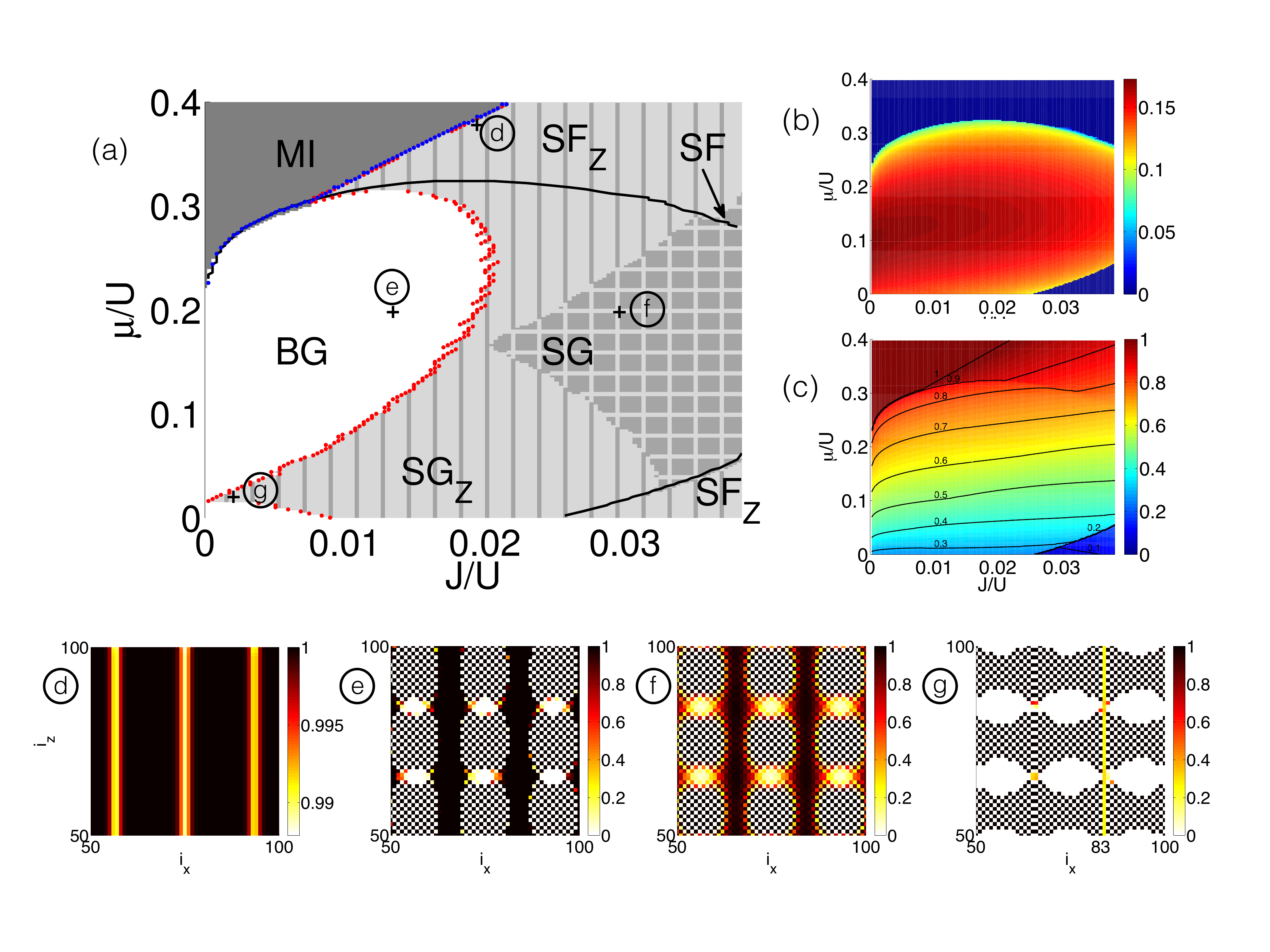}\\
\caption{\label{PhaseDisordered1}(a) Phase diagram in the $J-\mu$ plane evaluated from Eq. \eqref{HLMF} using LMF with SF cluster analysis when the atoms are trapped in squared lattice with interparticle distance $a=\lambda_0/2$ and interacting with a cavity field along the $z$ direction with $\lambda=\lambda_0(1-\epsilon)$ with $\epsilon\simeq 0.05$. (a) Classification of the phases (according to Table \ref{Table:1}), (b) contour plot of $\langle \hat \Phi \rangle$ (b)  and (c) of the atomic density $\bar n$. In (a) the border of the MI region (blue dots) marks the appearance of SF sites. The border of the BG region (red, dots) marks the percolation of the SF cluster.  The black line delimits the region where the intracavity field is non-vanishing and the atomic density modulation exhibit quasi-periodic order supporting Bragg scattering into the cavity. The black crosses, which are labeled with the circled letters, mark the points chosen in Figure (d)-(g) for showing the local boson occupation number $\langle\O[i]{n}\rangle$ across the lattice. The points correspond to the parameters (d) $J/U=0.0194,\,\mu/U=0.378$, (e) $J/U=0.0134,\,\mu/U=0.198$, (f) $J/U=0.0296,\,\mu/U=0.198$ and (g) $J/U=0.0002,\,\mu/U=0.02$.}
\end{center}
\end{figure}
\end{widetext}

We now consider the case when $\lambda$ and $\lambda_0$ are incommensurate, so that the resulting Hamiltonian is aperiodic
and $\sum_jZ_j/K\simeq 0$. The calculations we present are for the choice $\lambda=\lambda_0(1-\epsilon)$, where 
$\epsilon_\lambda\simeq 0.05$. This specific case has been analysed by some of us in Ref. \cite{Habibian:2013} in one dimension, where we predicted, besides the
MI and SF phases, the appearance of BG phases where the atomic density is quasi-periodically modulated, so to allow for coherent scattering into the cavity.
In two dimensions mean field analysis identified relatively large regions of the phase diagram, where the phase is SS. This structural quasi-periodic order, in fact, is a Bragg grating which
maximizes the cavity-induced long-range interaction: the onsite density $\langle \hat n_i\rangle$ oscillates at the beating wave number $k_0-k$, such that when convoluted with
the amplitude $Z_i$ it gives rise to a nonvanishing value of $\langle\hat \Phi\rangle$ and thus of the field. Being the Hamiltonian aperiodic, the resulting phases exhibit the features of a glass. 

In this Section we analyse the resulting phase diagram in two dimensions by means of the LMF with SF cluster analysis for Hamiltonian \eqref{HLMF}. 
Figure \ref{PhaseDisordered1}(a) displays the phase diagram in the $J-\mu$ plane, which we have restricted to the region of parameters where the onsite density is between 0 and unity. The parameter region where Bragg order is found is delimited by the black line: within this border the intracavity photon number is nonvanishing, see subplot (b). This region exhibits a rich number of phases, which we classify as BG, SS, and SG$_z$: The BG phase is vertically separated by a SG$_z$ phase, which persists also at vanishing tunneling $J$. The SG$_z$ phase, in turn, is almost vertically splitted into two regions by the tip of SS phase, which then broadens as $J$ is increased and become a SF phase at the black line, where the intracavity photon number vanishes. Comparison with subplot (c), displaying the corresponding mean atomic density, shows that the intracavity photon number is maximum about $\bar n=0.5$, which is the value where the CDW is formed in the commensurate case. We further observe that at sufficiently low $J$ there seems to be a direct transition between MI and BG phase, analogous to the direct transition from MI to CDW in the commensurate case. Differing from the commensurate case, the density within the BG phase varies continuously with $\mu$, being the phase compressible.

The details of the density distribution give further insight. In the region where the intracavity photon number vanishes, the MI and the SF phases are separated by the SF$_z$ phase. This phase is characterized by SF stripes along $z$ induced by the transversal standing wave laser which is incommensurate with the confining potential. The density distribution is shown in subplot (d), the stripes are quasi-periodic along the laser propagation direction ($x$ axis). Moving along a line at constant density with $1>\bar n>0.9$ towards $J\to0$, the phase remains always SF$_z$. At lower values of the densities, $0.9>\bar n>0.4$, the SF$_z$ has a transition to a SG$_z$ phase, namely, the SF stripes start to shrink while the density becomes modulated exhibiting Bragg order, as shown in subplot (f). Further decreasing $J$ shrinks the width of the SF stripes until they all split up into isolated islands, as visible in subplot (e): This signals the BG phase.

For densities $\bar n\sim 0.4$ the phase is SG down to zero tunneling, the structure is shown in subplot (g). This density separates two BG regions with different features of the structure form factor, corresponding to the two different patterns one can form so to match Bragg order. Roughly speaking, the patterns in the upper BG lobe corresponds to the one of the lower one by substituting particles with holes. The density separating the two regions cannot match both conditions simultaneously, unless particles can percolate along one direction, which could be a reason why the phase is SG$_z$ down to zero tunneling.
A similar behaviour is observed for the SS region. This region separates two SG$_z$ regions which are qualitatively different. In this case, this becomes visible in the SF cluster analysis by the structure of the patterns of MI and SF clusters. In both cases the different phases are signaled by the appearance or disappearance of an additional component of the structure form factor. Differing from the one dimensional case of Ref. \cite{Habibian:2013}, thus, the formation of an intracavity field sustains a variety of phases which can exhibit superfluidity. In addition to the MF study of Ref. \cite{Habibian:2013}, moreover, the LMF with SF cluster analysis further reveals that the selforganized phase exhibits different orders, such as SG$_z$ and SG, and also qualitatively different patterns. The onsite repulsion plays an important role, setting an additional constrain to achieving the Bragg order, and can force the system to remain a SG.

We have further analysed the form of the phase diagram for different incommensurate cases taking $\lambda=\lambda_0/\ell +\epsilon_{\lambda}$. For $\ell$ odd, the phase diagram present the same phases as for $\ell=1$, the main difference is in the fractional densities which constrain the formation of a BG phase. For $\ell$ even we also observe the formation of BG phases at the border of the MI phase, which then become SG at the interface: The system tends to form Bragg ordered structured at low densities. In general, this behaviour shows that the long-range interaction favours the formation of quasi-periodic structures supporting scattering into the cavity mode.  

\section{Conclusion}
\label{Sec:Conclusions}

To conclude we have used the local mean field approach with SF cluster analysis to calculate the phase diagrams for the extended 
Bose-Hubbard model with cavity mediated long range interactions. When the cavity wavelength $\lambda$ is
commensurate with the one of the external periodic potential $\lambda_0$, the phase diagram is 
similar to the phase diagram of the extended Bose-Hubbard model with
repulsive nearest neighbour interactions. For $\lambda=\lambda_0$, in particular, it exhibits MI lobes around half-integer values
of the ratio between chemical potential and onsite repulsion, $\mu/U$, and incompressible CDW checkerboard lobes 
around integer values. Each tip of the CDW lobes is covered with
a small super-solid (SS) region. We expect the phase diagram to be 
qualitatively identical in the experimental two-dimensional setup,
sketched in Fig. \ref{Fig:1}, and also for higher dimensions ($d>2$). 
In one space dimension LMF is a rather crude approximation, but still we expect 
similar features of the phase diagram to occur \cite{Batr:1995,Scal:1995,Fazio:2011}.

In the incommensurate case, for $\lambda=\lambda_0(1-\epsilon)$
with $\epsilon\ll 1$ our LMF calculation combined with
a SF cluster analysis \cite{Niederle:2013,Niederle:2015} reveals a much richer
phase diagram than predicted by conventional mean-field theory \cite{Habibian:2013}:
In addition to MI, SF$_z$, and SF phases, we could identify also an isotropic and
striped superglass (SG) phase, which 
is characterised by an aperiodic (glassy) density modulation
and isotropic or striped superfluid (off-diagonal) order. The striped SG regions are characterized
by superfluid stripes in $z$ direction, implying 
off-diagonal order (i.e., phase coherence) along these stripes. In an experimental setup, such as the one of Ref. \cite{Baumann:2010}, one would expect an exponentially small 
overlap between the wave function of the different stripes leading to extremely small but non-vanishing phase coherence  also in the perpendicular direction. 

We note that similar striped superfluid phases have also been identified 
in the extended Bose-Hubbard model with nearest and next-nearest neighbor
repulsion \cite{Batr:1995,Scal:1995}.  Moreover, we expect that SG phase shall also occur in the two-dimensional Bose-Hubbard
model with an aperiodically modulated chemical potential,
when the modulation wave vector is only slightly different
from the lattice constant. 

We finally remark that, even though superior to a conventional MF approach, the LMF method underestimates the effect of quantum fluctuations that can change the critical properties, as is well known in the disordered case \cite{BH-dis-1,BH-dis-2,BH-dis-3,BH-dis-4,BH-dis-5,BH-dis-6,Laflorangie:2015}. Therefore it would be desirable to check our predictions for the commensurate as well as for the incommensurate case by means of quantum Monte Carlo simulations.

\acknowledgments
The authors are grateful to Hessam Habibian, Nishant Dogra, Andr\'e Winter, Benjamin Bogner, Eugene Demler and Rosario Fazio for discussions and helpful comments, and the German Research Foundation (DACH project: "Quantum crystals of photons and atoms") for financial support.

\end{document}